\documentclass[prd, aps, superscriptaddress, preprintnumbers, twocolumn, floatfix, nofootinbib]{revtex4}

\usepackage{amsfonts}
\usepackage{amsmath}
\usepackage{amssymb}
\usepackage{bm}
\usepackage{dcolumn}
\usepackage{graphicx}   
\usepackage[latin1]{inputenc}
\usepackage{latexsym}
\usepackage{rotating}
\usepackage{hyperref}
\usepackage{graphicx}
\usepackage{color}

\newcommand\be{\begin{equation}}
\newcommand\ba{\begin{eqnarray}}
\newcommand\ee{\end{equation}}
\newcommand\ea{\end{eqnarray}}

\begin{document}

\title {Dark Energy, Dark Matter and Baryogenesis from a Model of a Complex Axion Field}

\author{Robert Brandenberger}
\email{rhb@physics.mcgill.ca}
\affiliation{Department of Physics, McGill University, Montr\'{e}al, QC, H3A 2T8, Canada and
Institute of Theoretical Physics, ETH Z\"urich, CH-8093 Z\"urich, Switzerland}

\author{J\"urg Fr\"ohlich}
\email{juerg@phys.ethz.ch}
\affiliation{Institute of Theoretical Physics, ETH Z\"urich, CH-8093 Z\"urich, Switzerland}

\date{\today}

\begin{abstract}

We introduce and study a model designed to simultaneously shed light on the mysteries connected with Baryogenesis, Dark Matter and Dark Energy. The model describes a self-interacting complex axion field whose imaginary part, a pseudo-scalar axion, couples to the instanton density of gauge fields including the hypermagnetic field. This coupling may give rise to baryogenesis in the early universe. After tracing out the gauge and matter degrees of freedom, a non-trivial effective potential for the angular component of the axion field is obtained. It is proposed that oscillations of this component around a minimum of its effective potential can be interpreted as Dark Matter. The absolute value of the axion field rolls slowly towards $0$. At late times, it can give rise to Dark Energy. 

\end{abstract}

\maketitle

\section{Introduction} 
\label{sec:intro}

In this letter we propose and describe some specific theoretical ideas on the origin of Dark Matter, Dark Energy and baryogenesis. The dark sector is known to make up about $95\%$ of the energy density of the universe. Roughly $70\%$ of the total energy density corresponds to Dark Energy, while approximately $25\%$ originates in Dark Matter; see, e.g., \cite{Data}. Dark Matter has an equation of state given by $w \simeq 0$, where $w$ is the ratio between the universe's pressure, $p$, and energy density, $\rho$, whereas the equation of state of Dark Energy is known to be $w \simeq -1$. 

A conventional candidate for a Dark Matter particle is a WIMP ($=$weakly interacting massive particle, see \cite{WIMP} for a review), and Dark Energy is usually described by a small cosmological constant. However, these simple descriptions of Dark Matter and Dark Energy appear to meet with increasing difficulties. The WIMP model of Dark Matter faces the problem that WIMP's have not been observed in any direct detection experiments, which rules out part of the preferred parameter space \cite{noWIMP}. For what concerns Dark Energy, there is increasing evidence that a positive cosmological constant cannot appear in current theories of quantum gravity \cite{swamp, Dvali}. There are thus good reasons -- see e.g. the discussion in \cite{swamp2} -- to imagine that Dark Energy is described by dynamical degrees of freedom, such as the slowly rolling scalar field introduced in Quintessence models \cite{Quintessence}. Oscillating pseudo-scalar fields with a small mass, such as an axion field, have long been envisaged as candidate degrees of freedom describing Dark Matter; see, e.g., \cite{axionDM} for a review. 

From a theorist's point of view it would be attractive if Dark Matter and Dark Energy turned out to have a common origin. This is the theme developed in this Letter. We introduce a model of a complex scalar field, 
$\zeta = e^{-(\varphi + i \theta)/f}$, whose radial component, $\varphi$, gives rise to Dark Energy, 
while the angular component, 
$\theta$, is supposed to describe Dark Matter, and $f\approx m_{pl}$ is a constant of nature rendering
$(\varphi + i \theta)/f$ dimensionless; see \cite{us-early} for earlier attempts, and \cite{review} for a review of 
various unified dark sector models. The self-interaction potential of the field $\zeta$ is assumed to be proportional to 
$\bar{\zeta} \zeta = e^{-2\varphi/f}$; (it is exponential in the scalar field $\varphi$, as in Quintessence 
models \cite{Quintessence}). We further assume that the imaginary part of $\zeta$ is coupled to the instanton 
density of some gauge fields in a way analogous to how the QCD axion is coupled to the color gauge field. When 
tracing out the gauge and matter degrees of freedom, this coupling generates a potential for the angular component, 
$\theta$, of $\zeta$, which gives rise to oscillations of the 
pseudo-scalar axion field $\theta$ around a minimum. These oscillations are a source of Dark Matter. The radial part, 
$\varphi$, of $\zeta$ slowly grows towards very large values, and hence the potential, $\propto e^{-2\varphi/f}$, 
slowly approaches 0. This potential is a source of (dynamical) Dark Energy.

An intriguing feature of our model is that it also naturally incorporates a mechanism for baryogenesis. The imaginary part of 
$\partial_{\mu} \zeta$ can be coupled to the anomalous axial baryon current, $j^{\mu}_{B}$. During an era when the time derivative of $\Im \zeta$ (or of $\theta$) has a fixed sign this coupling gives rise to a matter-antimatter asymmetry; see also \cite{Cohen, Kamada-Long, us}.

The organization of this Letter is as follows. In the next section we introduce the model studied afterwards. In Section 3 we discuss constraints on the parameters of the model and show that we can satisfy all the known constraints derived from the requirement that one wants to obtain the right amount of Dark Matter and Dark Energy. We discuss baryogenesis in Sections 4 and 5. Section 6 contains some conclusions. In an Appendix we discuss possible roots of our model in more fundamental physical theories.

Throughout this paper we employ natural units in which the speed of light, Planck's constant and Boltzmann's constant are set to 1. The cosmological scale factor appearing in the equations of the Friedman-Lema\^{i}tre universe is denoted by $a(t)$, where $t$ denotes time. The radiation temperature, $T$, is related to time $t$ via the Friedmann equation and the Stefan-Boltzmann law. The Hubble expansion rate is denoted by $H(t)$, and the Planck mass by $m_{pl}$.  There are various times which play a role in our analysis: The current time is denoted by $t_0$, the time of equal matter and radiation is $t_{eq}$, and the time after which the dynamics of the universe starts to be described by our model is denoted by $t_c$. The corresponding radiation temperatures are $T_0$, $T_{eq}$ and $T_c$, respectively. We sometimes express time in terms of the cosmological redshift, $z$. The redshift at time $t$ is defined by
\be
z(t) + 1 \, \equiv \, \frac{a(t_0)}{a(t)} \, .
\ee

\section{The Model} \label{model}

As announced in the Introduction, the model studied in this Letter describes a complex scalar field
\be
\zeta \, = \, e^{ - (\varphi + i \theta) / f} \, ,
\ee
where $\varphi$ is a real scalar field, called the ``radial component'' of $\zeta$, $\theta$ is a real pseudo-scalar axion field, called ``angular component'' of $\zeta$, and $f$ is the field range over which the potential, $\bar{\zeta}\zeta$, of $\zeta$ varies appreciably. We introduce the one-form 
\begin{equation}\label{WZW}
j:=\zeta^{-1}d\zeta,\quad \text{i.e., }\,\, j_{\mu}=\zeta^{-1}\partial_{\mu} \zeta = -\partial_{\mu}(\varphi + i\theta)/f\,.
\end{equation}
Let $CS(G)$ denote the Chern-Simons 3-form of a gauge field $G$. 
An example of a plausible action functional is given by
\begin{align}\label{z-action}
S(\bar{\zeta}, \zeta, G):= \int &d^{4}x\sqrt{-g}\big(f^{2}\,\bar{j}_{\mu}\,j^{\mu} - \Lambda\, \bar{\zeta}\zeta\big)
 \nonumber \\
&- \lambda \int d\Theta \wedge CS(G)\,,
\end{align}
where $g$ is the determinant of the space-time metric (with components $g_{\mu \nu}$), $\Lambda$ is a constant of (mass) dimension 4, $\lambda$ is a dimensionless coupling constant, and
$$\Theta = \theta, \text{ or  }\,\Theta= \Im \zeta \,.$$
(Terms proportional to masses of matter degrees of freedom are neglected in \eqref{z-action}.)

After a phase transition at some temperature $T_c$, the non-abelian gauge degrees of freedom acquire a mass and are traced/integrated out. This yields an effective action for the field $\zeta$ of the form
\begin{align} \label{action}
S \, = \, \int d^4x \sqrt{-g}& \bigg\{ \frac{1}{2} \partial_{\mu} \varphi \partial^{\mu} \varphi
+  \frac{1}{2} \partial_{\mu} \theta \partial^{\mu} \theta - \nonumber \\
&- \Lambda e^{-2\varphi/f} - V(\varphi, \theta) + \dots \bigg\} \, ,
\end{align}
where $V(\varphi, \theta)= \mathcal{O}(\theta^{2})$, for $\theta \approx 0$, and where the dots stand for couplings of 
$\theta$ to the instanton density of the hypermagnetic $U(1)_{Y}$ gauge field. Choosing $\Theta= \Im \zeta$, one finds that, 
for small values of $\text{sin}(\theta/f)e^{-\varphi/f}$,
\be \label{pot}
V(\varphi, \theta) \, = \,\frac{1}{2}\mu^{4} \text{sin}^{2}(\theta/f) e^{-2\varphi/f}
\ee
where $\mu$ is some mass scale. Note that if $T_c$ is so large that $\varphi$ is negative, with $\vert \varphi \vert$ large enough, at the time of the phase transition, then this transition may be followed by some cosmological ``wetting transitions'', as studied in \cite{us}.

The equations of motion for the fields $\varphi$ and $\theta$, with $V$ as in \eqref{pot}, are given by
\ba
{\ddot{\varphi}} + 3 H {\dot{\varphi}} \, &=& \, 
\bigl[ \frac{2}{f} \Lambda + \frac{\mu^4}{f} {\rm{sin}}^2 \frac{\theta}{f} \bigr] e^{-2 \varphi / f} \label{phieq}\,, \\
{\ddot{\theta}} + 3 H {\dot{\theta}} \, &=& \, 
- \frac{\mu^4}{f} {\rm{sin}} (\frac{\theta}{f})   {\rm{cos}} (\frac{\theta}{f}) e^{-2 \varphi / f} \, , \label{thetaeq}
\ea
where terms involving the hypermagnetic gauge field have been neglected.

We propose to explore the possibility that $\varphi$ gives rise to dynamical Dark Energy, while oscillations of 
$\theta$ about the minimum of the potential $V$ are a source of Dark Matter. Besides the dark sector fields $\varphi$ and 
$\theta$, radiation contributes to the pressure and the energy density of the early universe. We assume that, after the phase transition, the contribution of radiation to the energy density dominates in the early universe. As a consequence, space is expanding, and the oscillations of $\theta$ are damped. This implies that, in equation (\ref{phieq}), the first term on the right side becomes the dominant term at late times. 

If one wants this model to predict the observed energy densities of Dark Energy and Dark Matter then the amplitudes of the two terms on the right side of (\ref{phieq}) must have roughly the same mean at redshifts close 
to $z = 2$, when Dark Energy starts to dominate. We assume that the initial value of $\varphi$ in the very early universe is negative, as is typically done in Quintessence models. We further assume that the potential for $\theta$ is generated at some early time corresponding to a temperature $T \approx T_c$, and that, at that time, the initial condition for $\theta$ is close to a local maximum of its potential. With these assumptions, we must consider three time periods in the  evolution of the universe predicted by our model: the late period when Dark Energy dominates; the intermediate era when Dark Matter dominates over Dark Energy; and the early epoch when $\theta$ is close to a local maximum of its potential and radiation dominates. 

The {\bf Dark Energy era}  is described, approximately, by an exact solution of the second order differential equation (\ref{phieq}), neglecting the second term in the parenthesis on the right side, which is given by
\be \label{sol}
\varphi(t) \, = f {\rm{ln}}(\beta t) \, ,
\ee
where $\beta$ is a constant that can be determined by inserting the ansatz (\ref{sol}) into (\ref{phieq}), with $H$ expressed in terms of the Friedmann equation
\be
H^2 \, = \, \frac{1}{3} m_{pl}^{-2} \bigl[ V + \frac{1}{2} {\dot{\varphi}}^2 \bigr] \, .
\ee
We obtain a quadratic equation for $\beta^2$. In the limit $f \gg m_{pl}$ the solution for $\beta$ is
\be
\beta^2 \, \simeq \frac{4}{3} \frac{\Lambda}{f^2} \bigl( \frac{m_{pl}}{f} \bigr)^2 \, ,
\ee
which yields an equation of state 
\be
w \, \simeq \, - 1 + \frac{4}{3} \bigl( \frac{m_{pl}}{f} \bigr)^2 \, .
\ee
In this limit, the same result can be obtained by means of the slow-roll approximation. As discussed in detail in  \cite{exppot}, this solution is a late time attractor.

We observe that the dependence of $\varphi$ on time
is small on a Hubble time scale. Thus, in Eq. \eqref{thetaeq}, we can assume $\varphi$ to be constant, namely equal to the value it has at the time $t_i$ when Dark Energy begins to dominate. In this approximation, (\ref{thetaeq}) becomes the equation of motion for a damped harmonic oscillator with frequency
\be \label{freq}
\omega \, \simeq \, \frac{\mu^2}{f} e^{- \varphi(t_i) / f} \, .
\ee
A self-consistency condition for the validity of this approximation is that the frequency $\omega$ must be large as compared to the Hubble expansion rate. As we will see later, this condition is satisfied. Since the potential for $\theta$ is quadratic in the vicinity of its minimum, the equation of state of the degrees of freedom corresponding to the field $\theta$ corresponds to that of pressureless Dark Matter. The amplitude, ${\cal{A}}(t)$, of the oscillations of $\theta(t)$ decreases as
\be
{\cal{A}}(t) \, \sim \, a(t)^{-3/2} \, \sim \, T(t)^{3/2} \, .
\ee

Next, we turn to the analysis of the evolution of the fields in the {\bf intermediate era}: The field $\theta$ exhibits damped oscillations, as in the Dark Energy phase, but the $\theta$-dependent term dominates the right side of the equation of motion (\ref{phieq}), which then takes the form
\be \label{phieq2}
{\ddot{\varphi}} + 3 H {\dot{\varphi}} \, \simeq \, 
 \frac{\mu^4}{f} {\rm{sin}}^2 \frac{\theta}{f} e^{-2 \varphi / f}\, .
 \ee
 We set $e^{-2 \varphi / f} = 1$; later, we verify that this assumption is self-consistent. Furthermore, we replace the 
 ${\rm{sin}}^2$-term by its time average and make the small-angle approximation, with the source term quadratic in the amplitude. When the time dependence of the amplitude of $\theta$ is inserted, Eq. (\ref{phieq2}) becomes a first-order inhomogeneous differential equation for $\chi \equiv {\dot{\varphi}}$; namely
 \be \label{chieq}
 {\dot{\chi}} + \frac{2}{t} \chi \, = \, \frac{1}{2f} {\cal{A}}^2(t_{eq}) \mu^4 \bigl( \frac{t_{eq}}{t} \bigr)^2 \, ,
 \ee
 where $t_{eq}$ is the time of equal matter and radiation, and we have inserted the formula for $H$ during the matter-dominated era. The solution of this equation is
 \be
 \chi(t) \, = \, \frac{\alpha}{t} \, ,
 \ee
 with
 \be
 \alpha \, = \, \frac{\mu^4}{f^3} {\cal{A}}^2(t_{eq}) t_{eq}^2 \, ,
 \ee
 which implies that
 \be \label{result1}
 \varphi(t) \, = \, \alpha\, {\rm{ln}} \bigl( \frac{t}{t_{eq}} \bigr) + \varphi(t_{eq}) \, .
 \ee
 It is easy to check that $\alpha / f \ll 1$. Hence, the time dependence of $\varphi$ is
 negligible in this phase.
 
 Note that formula (\ref{result1}) is valid for $t > t_{eq}$. A similar analysis of the equations of motion applies for times before recombination. All that changes is the coefficient of the Hubble damping term in (\ref{phieq2}), as well as the time dependence of the amplitude of the oscillations of $\theta$. With approximations identical to those made above, the equation of motion for $\chi$ becomes
\be
   {\dot{\chi}} + \frac{3}{2t} \chi \, = \, \frac{1}{2f} {\cal{A}}^2(t_{eq}) \mu^2 \bigl( \frac{t_{eq}}{t} \bigr)^{3/2} \, ,
\ee
 whose solution implies that
 \be
 \varphi(t) \, = \beta t^{1/2} \, + {\rm{const}} ,
 \ee
 with 
 \be
 \beta \, = \, \frac{1}{f^3} {\cal{A}}^2(t_{eq}) t_{eq}^{3/2} \mu^4 \, .
 \ee
 Given the parameter values discussed below, it is easy to check that $\varphi$ varies very slowly as a function of time.   
  
 We now discuss the evolution of the fields $\varphi$ and $\theta$ during the {\bf initial era}, right after the phase transition when the potential $V(\varphi, \theta)$ can be used to study the evolution. As mentioned at the beginning of this section, we assume that $\theta$ starts close to a local maximum of its potential, here taken to be $\theta =  - (\pi f) / 2$. We are interested in analyzing the growth of the deviation, $\Delta \theta$, of $\theta$ from its value at the local maximum of the potential, which is given by
 \be
 \theta \, = \, - \frac{\pi f}{2} + \Delta \theta \, .
 \ee
 For small values of $\Delta \theta$, its equation of motion can be approximated by
 \be
  (\Delta \theta)^{\cdot\cdot} + \frac{3}{2t} (\Delta \theta)^{\cdot} \, \simeq \, \frac{\mu^4}{f^2} e^{- 2 \varphi / f} \Delta \theta \,.
 \ee
 The approximate behavior of $\varphi$ during this time period is described by the equation
 \be\label{EOM}
 {\ddot{\varphi}} + 3H {\dot{\varphi}} \, \simeq \, \frac{\mu^4}{f} e^{- 2 \varphi / f} \, .
 \ee
 This equation has the exact solution
 \be \label{Iphiev}
 \varphi(t) \, = \, \varphi_0 {\rm{ln}} (\gamma t)  \, ,
 \ee
 where $\varphi_0 = f$ is chosen to ensure that all terms in the equation have the same time dependence, 
 and where $\gamma$ is chosen so as to make the coefficients match. In the radiation phase we find that
 \be \label{Itimescale}
 \gamma \, = \, \sqrt{2} \frac{\mu^2}{f} \, .
 \ee
 Thus,
 \be \label{phisolution}
 e^{- 2 \varphi / f} \, = \, \frac{f^2}{2 \mu^4}\frac{1}{t^2} \, .
 \ee
  Inserting (\ref{phisolution}) into the equation of motion for $\Delta \theta$  we obtain
 \be
  (\Delta \theta)^{\cdot\cdot} + \frac{3}{2t} (\Delta \theta)^{\cdot} \, = \, \frac{1}{2 t^2} \Delta \theta \, 
  \ee
 where, here, $\Delta \theta$ is the deviation
of $\theta$ from the minimum of its potential. The dominant solution is
  \be \label{result3}
  \Delta \theta (t) \, \sim \, \frac{t^{1/2}}{t_c^{1/2}} \Delta \theta(t_c) \, ,
  \ee
where $t_c$ is the initial time. We will use these solutions for $\Delta \theta$ and $\varphi$ in the sections on baryogenesis.

We have corroborated the approximate analytical analysis presented above by a numerical study: the
system of three coupled differential equations consisting of (\ref{phieq}) and (\ref{thetaeq}) for the fields $\varphi$ 
and $\theta$ and the Friedmann equation
\be
H^2 \, = \, \frac{m_{pl}^{-2}}{3} \bigl[\frac{1}{2} {\dot{\varphi}}^2 + \frac{1}{2} {\dot{\theta}}^2 +  \Lambda e^{-2\varphi/f} + V(\varphi, \theta) \bigr] \, .
\ee
for the Hubble parameter $H(t)$, has been solved numerically.
In the simulations, all quantities are expressed in Planck units, including time, the dimensionless time being $\tau = m_{pl}\,t$. In 
Figure 1, the evolution of $\Delta \theta$ as a function of time is displayed. Figure 2 shows the evolution of $\varphi$ 
and of $\Delta \theta$ as functions of time. Figure 3 displays the time evolution of the parameter $w = p / \rho$ appearing in the total equation of state; and Figure 4 shows how the ratio of the contributions of Dark Matter and Dark Energy to the potential energy evolves.
In our simulation, we have chosen the parameter values $f = 2 m_{pl}$, $\mu = 5$ and $\Lambda = 10^{-8}$. The initial conditions have been set to be $\varphi = 0$, $\varphi^{\prime} = 1$, the prime denoting the derivative with respect to $\tau$, 
and $\theta$ displaced from the local maximum at $- \pi / 2$ by $\Delta \theta = 10^{-2}$, with $\dot{\theta}=0$.

The figures show that there are smooth transitions between the three epochs described in the text - the initial era where 
$\theta$ starts to slowly roll from a value very close to the one corresponding to the local maximum of the potential, the intermediate era when $\theta$ oscillates about the minimum of its potential, yielding an epoch of Dark-Matter domination, and - after the oscillations have redshifted sufficiently - the onset of the Dark Energy era when the ratio, $w$, of pressure to energy density approaches a negative constant.

\begin{figure}[h!]
\includegraphics[width=\hsize]{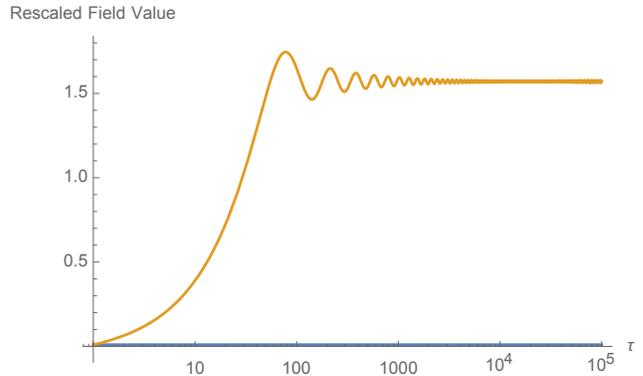}
\caption{Time evolution of the displacement, $\Delta \theta$, of the field $\theta$ from its value at the local maximum, for the parameter values and initial conditions chosen in the text. The field and time are in Planck units. After a time period of slow rolling, 
$\theta$ begins to oscillate about the minimum of its potential. The amplitude of oscillation is damped by the cosmological expansion. } 
\end{figure}

\begin{figure}[h!]
\includegraphics[width=\hsize]{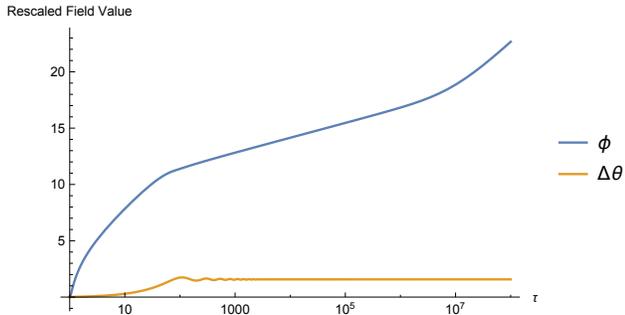}
\caption{Time evolution of both $\varphi$ and $\Delta \theta$.}
\end{figure}

\begin{figure}[h!]
\includegraphics[width=\hsize]{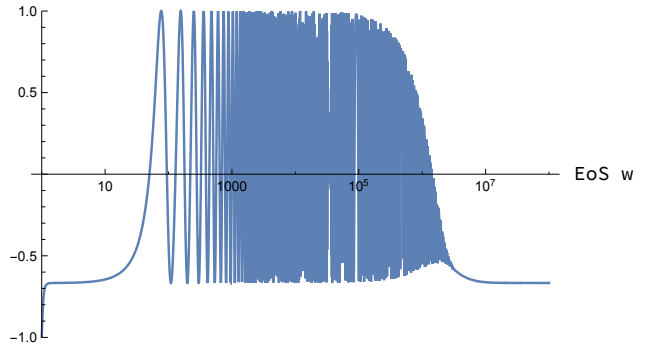}
\caption{Time evolution of the (total) equation-of-state parameter, $w$. This simulation does not take into account any radiation. Hence, initially, $w$ is negative, since the potential energy dominates over the kinetic energy. Once $\theta$ begins to oscillate about the minimum of its potential, the time average of $w$ has the value typical of Dark Matter ($w \sim 0$). Eventually, the energy stored in the oscillations of $\theta$ has redshifted sufficiently for the $\Lambda$ term in the potential to start to dominate. This signals the onset of the Dark Energy phase.}
\end{figure}

\begin{figure}[h!]
 \includegraphics[width=\hsize]{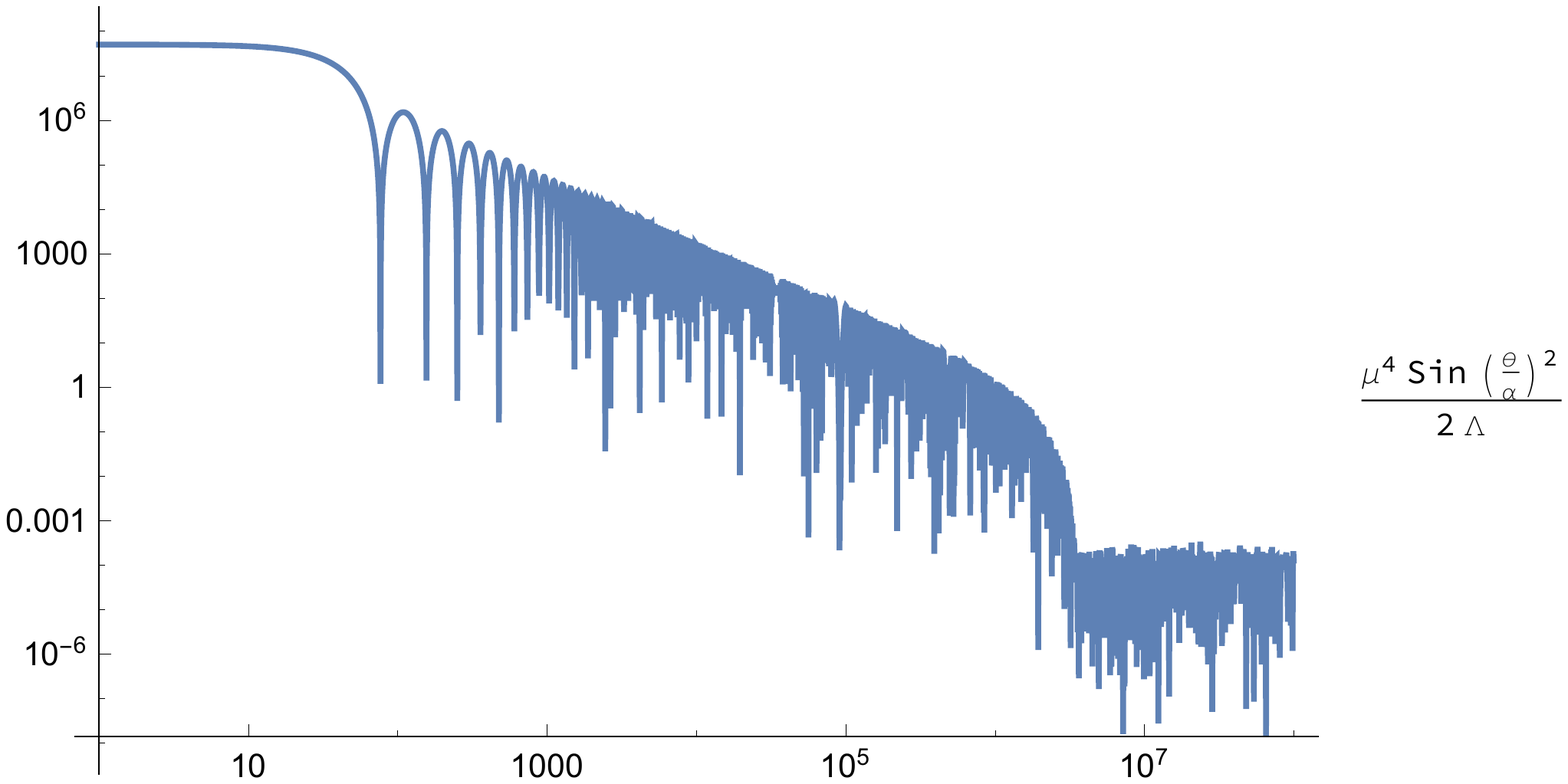}
 \caption{The ratio of the potential energy contributed by $V(\varphi, \theta)$ and the one contributed by the $\Lambda$ term. When this ratio drops below 1 the Dark Energy phase sets in.}
\end{figure}

The parameters in our numerical study have not been given realistic values, but have been chosen so as to facilitate the implementation of the numerics.

\section{Physical Constraints on Parameter Values}

There are four free parameters in our model, namely $f, \Lambda, \mu$ and $T_c$. We propose to estimate the values they must be given for our scenario to work. We start by recalling the various times involved in our analysis: $t_0$, the present time; $t_i$, the time when the Dark Energy era begins; and $t_c$, the time when the phase transition 
generating the potential $V$ for $\theta$ occurs. In the following we are only interested in the order of magnitude of 
the different terms appearing in our equations.

We note that, in order for the equation of state of $\varphi$ to correspond to the one of Dark Energy, we must impose the condition that $f \geq m_{pl}$, which is well-known in Quintessence models with exponential potentials. In the following we set $f = m_{pl}$ to simplify our estimates of the remaining parameters.

The first condition is that, at late times, the field $\varphi$ contributes the correct amount to the energy density of the universe to explain the currently observed Dark Energy. Since, in the Dark Energy era, the first term in the potential appearing in the action functional \eqref{action} dominates over the second one, the condition reads
\be \label{cond1}
\Lambda e^{- 2 \varphi(t_0) / f} \, \sim \, T_0^4 z_{eq} \, ,
\ee
where $T_0$ is the current temperature of radiation, and the factor $z_{eq}$ is the redshift at the time of equal matter and radiation. Its appearance in (\ref{cond1}) expresses the fact that the total energy density, today, is larger than the radiation energy density by that factor.

A condition on the mass scale $\mu$ is derived by demanding that, at the present time, the oscillations of $\theta$ 
yield the correct dark matter density. This condition reads
\be \label{cond2}
\mu^4 \frac{{\cal{A}}^2(T_0)}{f^2} e^{- 2 \varphi(t_0) / f} \, \sim \, T_0^4 z_{eq} \,.
\ee
In the rather rough estimates described here we are taking the contributions of Dark Energy and of Dark Matter to the current energy density of the universe to be the same.

As argued in the previous section, the value of $\varphi$, today, is close to $f$. This allows us to neglect the exponential factors in (\ref{cond1}) and (\ref{cond2}). Then (\ref{cond1}) becomes
\be \label{cond1b}
\Lambda \, \sim \, T_0^4 z_{eq} \, .
\ee
Using the fact that the temperature dependence of ${\cal{A}}$ is $\propto T^{3/2}$, and assuming that the initial amplitude is of the order of $f$, the second condition (\ref{cond2}) becomes
\be \label{cond2b}
\mu^4 \bigl( \frac{T_0}{T_c} \bigr)^3\, \sim \, T_0^4 z_{eq} \, .
\ee
Note that conditions (\ref{cond1b}) and (\ref{cond2b}) are similar to the tunings required in every known dynamical dark sector model: there is no explanation of the fact that Dark Energy, Dark Matter and visible matter yield comparable contributions (within one order of magnitude) to the total energy density of the universe just at the present time. It is important to check that, besides the tuning conditions that guarantee that this fact is properly reproduced by our model, no further fine-tuning of parameter values is required.

The value of the mass parameter $\mu$ determines the mass, $m_{DM}$, of Dark Matter modes, because this mass is given by the frequency (\ref{freq}) of oscillations of $\theta$. Setting the exponential factor in Eq. \eqref{freq} to 1, as above, we obtain
\be \label{cond3}
m_{DM} \, \sim \, \frac{\mu^2}{f} \, .
\ee
If we fix the Dark Matter mass by setting
\be \label{cond4}
m_{DM} \, \equiv \, m_a 1{\rm eV} \, ,
\ee
where $m_a < 1$ is a dimensionless number, then the value of $\mu$ is determined by $m_a$. Eq. (\ref{cond2b}) then determines $T_c$. We find that
\ba
\mu \, &\sim & \, m_a^{1/2} 10^5 {\rm GeV} \, \\
T_c \, &\sim & \, m_a 10^{14} {\rm GeV} \, .
\ea
In order to end up with a Dark Matter mass in the range of a typical axion mass $m \sim 1{\rm eV}$, the scales $\mu$ and $T_c$ are related to physics at very high energy scales. Yet, to obtain a mass $m_{DM} \sim 10^{-20} {\rm eV}$ corresponding to ultralight Dark Matter, the values of $\mu$ and $T_c$ must be in the range $\mu \sim T_c \sim 10^{4} {\rm{eV}}$.\\

\section{Baryogenesis in the Early Universe}

It is natural to assume that the imaginary part of the gradient of the complex scalar field $\zeta$ is coupled to
the baryon current $j_B^{\mu}$ by the term 
\be \label{deltaL}
\delta {\cal L} \, = \, {\tilde{\alpha}} \partial_{\mu} \Im \zeta \,j_B^{\mu}\, ,
\ee
where ${\tilde{\alpha}}$ is a dimensionless coupling constant. The presence of this term in the Lagrangian can be motivated by comparing it with the last term in \eqref{z-action} and recalling the chiral anomaly for the baryon current
\be
\partial_{\mu} j_B^{\mu} \, \sim \, \frac{g^2}{16 \pi^2} F \wedge F \, ,
\ee
where the masses of quarks are neglected, and where $F$ is a non-Abelian gauge field coupling to baryon number. 
The field $F$ could be the $SU(2)$ gauge field of the electroweak interactions; but we leave the question 
what the physical nature of $F$ is open; it is natural to suppose that it is the gauge field, previously 
denoted by $G$, that generates the potential for $\theta$ discussed at beginning of this paper.

Following \cite{Cohen}, we note that, during an era when $\Im z$ is rolling uniformly, the above interaction term 
generates a chemical potential, $\mu_B$, conjugate to baryon number
\be \label{chem}
\mu_B \, = \, {\tilde{\alpha}} (\Im \zeta)^{\cdot} \, = \, 
\frac{{\tilde{\alpha}}}{f} \bigl[ {\dot{\theta}} - \frac{{\dot{\varphi}}}{f}
 \theta \bigr] e^{- \varphi / f} \, .
\ee
In our cosmological scenario, $\Im z$ is rolling uniformly in the initial era, right after the phase transition generating
 the effective potential of $\theta$. Thus, we need to estimate the value of $\mu_B$ at these early times, making use of the results 
 found in (\ref{phisolution}) and (\ref{result3}). We find that
\be
\mu_B \, \sim \, {\tilde{\alpha}} \frac{T^4}{\mu^2 m_{pl}} \, ,
\ee
where we have used the Friedmann equation to express time $t$ in terms of the temperature $T$.

As long as baryon-number violating interactions involve degrees of freedom that are in thermal equilibrium during the early phase in the evolution of the universe we are considering here, a chemical potential $\mu_{B}$ corresponds to a baryon number density, $n_B$, of the order of
\be
n_B \, \sim \, \mu_B T^2 \, ,
\ee
and the induced baryon-number density-to-entropy ratio is found to be given by
\be
\frac{n_B}{s} \, \sim \, {\tilde{\alpha}} \bigl( \frac{T}{\mu} \bigr)^2 \frac{T}{m_{pl}} \, ,
\ee
which is to be evaluated for values of the temperature $T$ corresponding to the initial period of the field evolution, i.e., for a value of $T$ of the order of  $T_c$. 

For the baryogenesis scenario described here to work baryon-number violating processes must be in local thermal equilibrium when the temperature of the universe is close to $T_c$. Moreover, the era of slow rolling of $\theta$ and 
$\varphi$ must last at least a Hubble time in order for local thermal equilibrium to be established, which is what might justify introducing the chemical potential $\mu_B$. It is easy to check that this latter condition is satisfied. The time scale of slow rolling can be read off from (\ref{Iphiev}) and is given by $\gamma^{-1}$, see (\ref{Itimescale}). Since $f > m_{pl}$, 
it follows that $\gamma^{-1} >  H^{-1}$. Furthemore, we have to require that the slow rolling of $\theta$ occur around 
the time of the electroweak phase transition, when a non-vanishing baryon number is generated. This implies that 
another condition for the mechanism described here to work is that $T_c \geq T_{EW}$, where $T_{EW}$ is 
the temperature of electroweak symmetry breaking. 

To conclude this section we note that the term (\ref{deltaL}) in the Lagrangian of our model violates baryon number conservation. One would therefore expect that baryogenesis also occurs out of thermal equilibrium, i.e., that the assumption of local thermal equilibrium during baryogenesis made above is not really necessary. We discuss a possible scenario in the next section.\\

\section{Baryogenesis from Hypermagnetic Helicity}

It is well known that the baryon current is anomalous (see \cite{anomaly} for the original articles on the chiral anomaly, \cite{BGrevs} for reviews of applications of the chiral anomaly to baryogenesis, and \cite{Juerg2} for an application of the chiral magnetic effect of electromagnetism to magnetic field generation).  In particular, the change in baryon number is proportional to the change of the hypermagnetic helicity \cite{hyper} (see \cite{Shaposh,Kamada, Amber}):
\be
\Delta N_B \, = \, C_y \frac{\alpha_y}{8\pi} \Delta {\cal{H}} \, ,
\ee
where $\alpha_y$ is the hypermagnetic fine structure constant and $C_y$ is a constant depending on the particle content of the model used to describe visible matter; (see, e.g., \cite{Amber} for values of $C_y$).

The variation of the density, $h$, of the hypermagnetic helicity in time is given by
\be \label{helicity}
{\dot h} \, = \, - 2 <E \cdot B> \, ,
\ee
where $E$ and $B$ are the electric and magnetic fields of hypermagnetism, and the angular brackets indicate spatial averaging. Here and in the following we neglect the expansion of the universe. At the end of this section 
we will comment on the effects caused by its expansion. In a regime where the time derivative of the electric field can be neglected the equations of magnetohydrodynamics imply that \cite{Moore, Juerg}
\be \label{inst}
E \cdot B \, = \, \frac{1}{\sigma} B \cdot (\nabla \wedge B) \, ,
\ee
where $\sigma$ is the conductivity, whose order of magnitude is given by the temperature, i.e.,
\be
\sigma \, \sim \, T \, .
\ee
The Fourier modes, $A_k$, of the hypermagnetic gauge field $A$ contribute to the spatial average of $E \cdot B$. As shown, e.g., in \cite{Amber}, the expression for the spatial average of the right side of \eqref{inst} is given by
\be
<B \cdot (\nabla \wedge B)> \, = \, \int_k \frac{d^3k }{(2\pi)^3}\,|k|^3 \bigl( |A_{k, +}|^2 - |A_{k, -}|^2 \bigr) \, ,
\ee
where the subscripts $+$ and $-$ indicate the helicities of the modes.

We assume that the field $\zeta$ also couples to the hypermagnetic instanton density via a term
\be
\delta {\cal{L}}_2 \, = \, \alpha \frac{\Im \zeta}{4} {\widetilde{Y}_{\mu\nu}} Y^{\mu\nu} \, ,
\ee
where $\alpha$ is a dimensionless coupling constant, $Y_{\mu\nu}$ is the field strength associated with $A$, and 
$\widetilde{Y}_{\mu\nu}$ is its dual; (this term arises from the one in \eqref{z-action} by integration by parts, setting 
$\Theta= \Im \zeta$). The equation of motion for $A_k$ then becomes \cite{hyper, Amber}
\be \label{Aeq}
{\ddot{A}_{k, \pm}} + \bigl( k^2 \pm \alpha k (\Im \zeta)^{\cdot} \bigr) A_{k, \pm} \, = \, 0 \, ,
\ee
As shown in \cite{Peloso}, the pseudoscalar field $\Im \zeta$ can induce growth of the helicity of the hypermagnetic field. As long as the time derivative of $\Im \zeta$ has a fixed sign, a property it has in our model during the initial epoch of evolution, then, for small values of $k$, one helicity mode is enhanced, while the other one exhibits damped oscillations. In the following we estimate the amplification of the \textit{growing} Fourier modes (the helicity label on the Fourier modes $A_k$ is now omitted).

For small values of $\theta$ we can approximate $\text{cos}(\theta)$ by $1$ and $\text{sin}(\theta)$ by $\theta$ and find that
\be \label{Imz}
{(\Im \zeta)^{\cdot}} \, = \, \frac{1}{f} e^{- \varphi / f} \bigl( {\dot{\theta}} - \dot{\varphi} \frac{\theta}{f} \bigr)\,.
\ee
Inserting (\ref{phisolution}) and (\ref{result3}) for $\varphi$ and $\theta$ in the initial epoch, we find that the two terms on the ride side of \eqref{Imz} coincide, up to a factor of 2. Hence, in (\ref{Aeq}), we can replace $(\Im \zeta)^{\cdot}$ by 
$|\zeta| \dot{\theta} / f$. The equation of motion for $A_k$ then becomes
\be \label{Aeq2}
{\ddot{A_k}} + \bigl( k^2 \pm \alpha k e^{- \varphi / f} \frac{\dot{\theta}}{f} \bigr) A_k\, = \, 0 \, .
\ee
An approximate solution of this equation is obtained by  assuming that
\be
\dot{\theta} \, \sim \, \frac{f}{t_c} \, .
\ee
We also approximate $e^{- \varphi / f}$ by the value it has at the beginning of rolling. Taking into account that $f \sim m_{pl}$ we get
\be \label{Aeq3}
{\ddot{A_k}} + \big[( k^2 - \alpha k \bigl( \frac{T_c}{\mu} \bigr)^2 \frac{1}{\tau} \bigr] A_k\, = \, 0 \, .
\ee
We define the ``critical wavenumber'' $k_c$ by
\be \label{kcrit}
k_c \, = \, \alpha \tau^{-1} \bigl( \frac{T_c}{\mu} \bigr)^2 \, .
\ee
We then find that modes with $\vert k \vert < k_c$ are exponentially amplified, whereas modes with $\vert k \vert \geq k_c$ oscillate with constant amplitude. 

The growth of the modes $A_{k}$, for $\vert k \vert < k_c$, is shut off by back-reaction: the energy density of the field 
quanta produced by the growth of the unstable Fourier modes of $A$ cannot exceed the one of radiation before non-linear effects become important. (The logic here is similar to the one used to explain the termination of the preheating instability \cite{TB} in reheating after inflation; see, e.g., \cite{Rouzbeh} for recent reviews). The energy density of the field  quanta of $A$ is given by
\be
\rho_A \, \sim \, \int d^3k\, k^2 A_k^2 \, ,
\ee
an integral dominated by the contribution of the integrand around $k \sim k_c$. The amplitude of the $A_k$-mode at times $t > t_c$, starting from vacuum initial conditions at time $t_c$, is given by
\be \label{growth}
A_k (t) \, = \, \frac{1}{\sqrt{2k}} e^{(\alpha k k_c)^{1/2} (t - t_c)} \, ,
\ee
where the origin of prefactor $(\sqrt{2k})^{-1}$ is explained by recalling that the harmonic oscillator $A_k$ has been starting in its ground state. Considering the growth rate of the $A_k$ modes described in (\ref{growth}), with $k=0$, we obtain that
\be
\rho_A \, \sim k_c^4 e^{2 \alpha^{1/2} k_c (t - t_c)} \, 
\ee
The time when the growth of helicity ends is determined, approximately, by equating the energy density $\rho_{A}$ with the energy density of degrees of freedom contributing to radiation, which is proportional to $ T^4$. Since $T \sim T_c$, the length, $\delta t$, of the time interval during which the helicity grows is given by
\be \label{cutoff}
e^{2 \alpha^{1/2} k_c \delta t} \, \sim \alpha^{-4} \bigl( \frac{\mu}{T_c} \bigr)^8 \bigl( \frac{m_{pl}}{T_c} \bigr)^4 \, ,
\ee
where we have used (\ref{kcrit}).

Having determined the duration, $\delta t$, of the period over which the hypermagnetic helicity grows, we return 
to (\ref{helicity}) with the purpose  of estimating the baryon number density, $\Delta n_B$, produced during that period. 
We find that
\be
\Delta n_B \, \sim \, C_y \frac{\alpha_y}{8\pi} \frac{\Delta t}{\sigma} k_c^5 \frac{1}{4 \pi^2} 
e^{2 \alpha^{1/2} k_c \delta t} \, .
\ee
Inserting the cutoff value (\ref{cutoff}) we see that the dependence on the mass parameter $\mu$ and on $T_c$ drops out of this expression, and we obtain a robust order-of-magnitude estimate
\be
{\Delta n_B} \, \sim \,  C_y \frac{\alpha_y}{8\pi} \alpha^{-1/2} \, .
\ee
Thus, we see that the mechanism sketched has the required efficiency to produce the observed baryon number to entropy ratio.

\section{Conclusions and Discussion} \label{conclusion}

In this Letter we have introduced and studied a model of a complex field $\zeta = e^{-(\varphi + i\theta)/f}$ 
describing the presence of plausible amounts of Dark Matter and Dark Energy in the universe. At late times, the energy density stored in the radial part, $e^{-\varphi/f}$, of $\zeta$ can be interpreted as Dark Energy. The gradient of the imaginary part of $\zeta$ is coupled  to the anomalous baryon current and hence to gauge degrees of freedom. After a phase transition, it acquires  a periodic effective potential generated by integrating out the matter and gauge degrees of freedom. The field $\theta$ will then eventually start to oscillate about the minimum of its potential at $\theta = 0$, with a frequency (rest mass) that decreases in time like  $e^{-\varphi/f}$. These oscillations yield light Dark Matter. Assuming that, after the phase transition, $\theta$ exhibits a slow roll starting from an initial value close to a local maximum of its potential then, during the period of slow roll and before the oscillations of $\theta$ set in, a non-vanishing baryon number can be generated. It follows that the model discussed in this Letter may apparently describe baryogenesis in the very early universe, Dark Matter at intermediate times, and Dark Energy at late times.

The present model should be compared with another model inspired by a scenario 
proposed in \cite{Rodrigo} that has been introduced in a previous paper \cite{us}. 
In the latter model, Dark Matter and Dark Energy are assumed to originate from the 
dynamics of a single real scalar field, $\varphi$. However, an additional scalar field must be introduced to 
trigger a phase transition, reminiscent of what is known as a ``wetting transition'', from a phase 
where $\varphi$ produces a high density of Dark Matter to a low-density phase describing Dark Energy. 
A substantial amount of fine-tuning of the parameters is necessary in order for this model to satisfy 
known model-building- and cosmological constraints. 
In addition, the model can only describe tiny masses of Dark Matter modes corresponding to 
ultralight Dark Matter. The model discussed in the present paper does \textit{not} require as 
much fine-tuning as that other model. Its parameters can be adjusted so as to describe a wide 
range of Dark Matter masses. An additional advantage of the new model, as compared to the 
one studied in \cite{us}, is that its degrees of freedom naturally couple to the anomalous baryon 
current and to the hypermagnetic gauge field, and hence it may also describe baryogenesis.

The action of the model studied in this Letter appears to satisfy  constraints on effective field theories derived from superstring theory; see \cite{swamp2}. But, like other models of dynamical Dark Energy, it does not shed any 
light on the ``coincidence problem'', namely on the question why Dark Energy is becoming dominant precisely at the present time.

\section*{Acknowledgement}
\noindent We thank Z. Wang for generating the figures shown in this paper, and R. Namba for collaboration during initial stages of this project. RB thanks the  Pauli Center and the Institutes of Theoretical Physics and of Particle- and Astrophysics of the ETH for hospitality. The research at McGill is supported, in part, by funds from NSERC and from the Canada Research Chair program. JF is a member of the NCCR SwissMAP, which is sponsored by the Swiss National Foundation.

\section*{Appendix}

In this appendix we speculate about possible roots of the model studied in this paper in fundamental theories of Nature.
Our model describes a complex scalar field $\zeta$, which we write as
\be
\zeta \, = \, e^{ - (\varphi + i \theta) / f} \, ,
\ee
where $f$ is a mass scale. The angular variable $\theta$ plays a role similar to the one of the axion in QCD (see e.g. \cite{Peccei} for a review of the coupling of the QCD axion to the QCD gauge fields). We take the action for $\zeta$, in the absence of any couplings to matter and gauge fields, to be given by 
\be \label{bare}
S_0 \, = \, \int d^4x \sqrt{-g} \bigl[ \frac{1}{2} f^2 |\zeta|^{-2} \partial_{\mu} {\bar \zeta} g^{\mu \nu} \partial_{\nu} \zeta - \Lambda |\zeta|^2 \bigr] \, ,
\ee
where $\Lambda$ is a constant of mass dimension 4.

Complex scalar fields similar to the field $\zeta$ are ubiquitous in effective field theories derived from 
superstring theory. An example encountered in string theory is the axion-dilaton field
\be
{\hat{\tau}} \, = \, - e^{-\Phi} + i C_0 \,, 
\ee
where $C_0$ is an axion that originates in the Ramond-Ramond zero form, and $\Phi$ is the dilaton; see, e.g., \cite{Baumann} for a review. At the classical level, the potential is flat in the axionic direction as a consequence of the usual shift symmetry. World sheet- or D-brane instanton effects break this continuous symmetry to a discrete symmetry and generate a potential for the axion $C_0 \equiv a$ of the form
\be \label{pot2}
V(a) \, \sim \, g \mu^4 {\rm sin}(a/f) \, ,
\ee
where $\mu$ is a constant of mass dimension 1, and the dimensionless coefficient $g$ is determined by the string coupling constant, i.e.,
\be
g \, \sim \, e^{- \Phi} \, .
\ee
This is one example of how the potential in Eq. (\ref{pot}) could arise.

String theories on space-times with six compactified dimensions also tend to give rise to complex scalar fields, with a self-interaction potential of the kind we are considering in this paper. To be specific, we think of an internal space given by a Calabi-Yau manifold. The ten-dimensional metric can then be written as (see, e.g., the discussion in Section 3 of \cite{Baumann})
\be
ds^2 \, = \, e^{-6u(x)} g_{\mu\nu}(x) dx^{\mu} dx^{\nu} + e^{2u(x)} {\tilde{g}}_{ab}(y)dy^ady^b,
\ee
where $x^{\mu}$ are coordinates of the four-dimensional space-time, and $y^a$ are coordinates of the internal Calabi-Yau manifold with metric ${\tilde{g}}_{ab}(y)$, which we assume to be fixed. The real scalar field $u(x)$ encodes the overall scale of the internal manifold. The field $u$ is related to the real part of a complex modulus field, $T$, namely
\be
{\cal{R}}T \, \equiv \, e^{4u} \, .
\ee
The imaginary part of $T$ arises from the dimensional reduction of the four-form potential. In the low energy (supergravity) limit, dimensional reduction of the ten-dimensional Ricci scalar in the Einstein-Hilbert action yields a canonical kinetic term in the effective action of the field $u(x)$. If the internal manifold has negative curvature, then $u$ acquires a positive potential given by
\be
V(u) \, \sim \, e^{8u}\,. 
\ee
(See also \cite{FG} for a derivation of a complex scalar field from compactification of extra dimensions.)

String compactifications in the presence of fluxes \cite{Witten,Becker} exhibit further fields that, after 
dimensional reduction to a four-dimensional space-time, are complex scalars with an axionic angular variable. 
For example, in type IIB string theory compactified on a six-dimensional Calabi-Yau manifold $X_6$, axion fields, $a$, with a potential of the form given in \eqref{pot2}, which are space-time pseudo-scalars, arise naturally. The mass scale $f$ is then given by \cite{Baumann}
\begin{equation}
\frac{f}{m_{pl}}  \, \sim \, \bigl( \frac{l_s}{L} \bigr)^2 \, ,
\end{equation}
where $L$ is a length scale characteristic of the internal manifold $X_6$, and $l_s$ is the string length.

In supersymmetric gauge theories, the gauge coupling constant, $g$, and the vacuum angle, $\Theta$ appear in the combination $ \frac{4\pi}{g^2} - \frac{i \Theta}{2 \pi}$\, , 
see \cite{Witten2}. One may imagine that $g$ and $\Theta$ are related to the expectation value of a dynamical complex scalar field, $\hat{\tau}=\varphi + i\theta\,$, 
where $\varphi$ is a real scalar field and $\theta$ is a pseudo-scalar axion field, with
\begin{equation}
\frac{ \langle \varphi \rangle}{f} =  \frac{4\pi}{g^{2}}, \qquad \frac{\langle \theta \rangle}{f} = \frac{\Theta}{2\pi}\,, 
\end{equation}
with $f \approx m_{pl}$, as above; see e.g. \cite{Bilal} for a review. The field $\zeta$ appearing in the model discussed in this paper could be the exponential of ${\hat{\tau}}$, i.e.,
\be
\zeta \, \equiv \, e^{-{\hat{\tau}}/f} \,.
\ee
One may then argue that the gradient of the imaginary part of $\zeta$ couples to the 
Chern-Simons three-form of some non-abelian gauge field or to the anomalous 
baryon current.\\

\end{document}